\documentclass[%
 reprint,
showpacs,showkeys,
 amsmath,amssymb,
 aps,
]{revtex4-1}
\usepackage{amsmath}
\usepackage{cases}
\usepackage{amssymb}
\usepackage{CJK}
\usepackage{graphicx}
\usepackage{dcolumn}
\usepackage{bm}
\usepackage{color}
\usepackage{ulem}
\usepackage[pagewise]{lineno}

\begin{document}


\title{Reexamining the "finite-size" effects in isobaric yield
ratios using a statistical abrasion-ablation model}

\author{C. W. Ma$^{1}$}\thanks{ Email: machunwang@126.com}\thanks{mobile:15836181225}
\author{S. S.Wang$^{1}$}
\author{H. L. Wei$^{1}$}
\author{Y. G. Ma$^{2}$}\thanks{ Email: ygma@sinap.ac.cn}
\affiliation{$^{1}$Department of Physics, Henan Normal University, Xinxiang 453007\\
$^{2}$Department of Nuclear Physics, Shanghai Institute of Applied Physics, \\
Chinese Academy of Sciences, Shanghai 201800
}
\date{\today}

\begin{abstract}
The "finite-size" effects in the isobaric yield ratio (IYR), which are shown in the
standard grand-canonical and canonical statistical ensembles (SGC/CSE) method,
is claimed to prevent obtaining the actual values of physical parameters. The
conclusion of SGC/CSE maybe questionable for neutron-rich nucleus induced reaction.
To investigate whether the IYR has "finite-size" effects, the IYR for the mirror nuclei [IYR(m)]
are reexamined using a modified statistical abrasion-ablation (SAA) model. It is
found when the projectile is not so neutron-rich, the IYR(m) depends on the isospin
of projectile, but the size dependence can not be excluded. In reactions induced
by the very neutron-rich projectiles, contrary results to those of the SGC/CSE models
are obtained, i.e., the dependence of the IYR(m) on the size and the isospin of the
projectile is weakened and disappears both in the SAA and the experimental results.

\end{abstract}

\pacs{25.70.Pq, 21.65.Cd, 25.70.Mn, 24.60.-k}
\keywords{finite effect, neutron-rich nucleus, isobaric method}
\maketitle

The yield of fragments  has been demonstrated as a powerful tool to study the
information of nuclear matter and temperature in heavy-ion collisions (HIC). For instance,
the scaling of the ratios between the isotopic yields in reactions of similar measurements, namely the
isoscaling, has been extensively used to study the symmetry energy of hot emitting
source in the framework of various theoretical models as well as in experiments \cite{HSXuPRL00,ChenZQisos10,ZhouPei11,Fang07JPG,YGMaiso05,Dorso06iso,Dorsoiso06finite,Ono03iso,MBT_EPJA06iso,TianisoscalT05CPL,Botv02iso}.
In addition, the kinetic energy spectra of light particles and the double yield ratio of light
fragments are taken as thermometers \cite{Alb,JSWangT05,Nato95T,YGMa05T,Wada97T,MaCW13ComTheoPhys,SuJPRC12T_spec,MaCW12PRCT}.

Using a modified Fisher model (MFM), which bases on the free energy \cite{ModelFisher2},
the isobaric yield ratio (IYR) is used to study the symmetry energy of the fragments
\cite{Huang10,MaCW12EPJAasym,MaCW12CPL06asym,MaCW12CPL09asym,MaCW13PRCisoSB}. The correlation
between the IYR and energy term which contributes to the free energy of the fragment is
constructed to extract the physical parameters associated with the mass of nucleus \cite{MaCW11PRC06}.
In the IYR method, the energy terms which only depend on mass cancel out, thus makes
it convenient to investigate the symmetry-energy term in the mass formula. However,
it is claimed that the IYR method, which utilizes only one reaction system,
does not provide cancelation or minimization of the effects associated with mass
and charge constraints due to the "finite-size" effects in the standard grand-canonical
and canonical statistical ensembles (SGC/CSE) methods \cite{Souza12finite}.

The linear behavior of the IYR for the mirror nuclei [IYR(m)], which is discussed in
Refs. \cite{Huang10,MaCW11PRC06,Souza12finite}, is as follows,
\begin{eqnarray}\label{lnRmirror}
IYR(m)=\mbox{ln}[R(I+2,I,A)]=\mbox{ln}[R(1,-1,A)]  \nonumber \\
\equiv\mbox{ln}[ Y(A,1)/Y(A,-1)]= [(\mu_n-\mu_p)+a_c\cdot x]/T,
\end{eqnarray}
where $I=N-Z$ is the neutron-excess, $a_c$ is the Coulomb-energy coefficient,
$x\equiv2(Z-1)/A^{1/3}$, $\mu_n$ and $\mu_p$ are the neutron
and proton chemical potentials, respectively. This equation suits for both the SGC/CSE and MFM theories.

Depending on the volume of the reaction system, it was claimed that the "finite-size" effects of the IYR(m)
in SGC/CSE prevent one from obtaining precise information on the
nuclear properties using the isobaric ratio method, and the "finite-size" effects
are negligible only for system sizes much larger than those actually formed \cite{Souza12finite}.
Actually, in the similar isoscaling method, though the finite size effect is
believed to be minimized from the sources with the fixed $N/Z$, it is still
obvious in the isoscaling in some models \cite{Dorsoiso06finite}.

The yields of the fragments strongly depends on the isospin, as well as the neutron
and proton density ($\rho_n$ and $\rho_p$) distributions of the projectile nucleus.
Actually, the $\rho_n$ and $\rho_p$ distributions are important inputs in the nuclear
reaction models. However, there is no such information in SGC/CSE \cite{Das05PR,Souza12finite}.
And temperature in SGC/CSE is set to a certain value, but in fact the temperature
has a wide distribution observed in light fragments and heavy fragments
\cite{MaCW13ComTheoPhys,SuJPRC12T_spec,MaCW12PRCT,TianisoscalT05CPL}. Different
temperature values have been used in similar theories \cite{Souza12finite,Chau76Temp,Tsang06BET}.
The yields of fragments in SGC/CSE are not shown in Ref. \cite{Souza12finite}.
Considering the density distributions of neutron-rich nucleus, the SGC/CSE is
not a good model to evaluate the yields of fragments in the reactions induced by
the neutron-rich nucleus using the generalized parameters. Thus the inference based on
the yields of fragments in SGC/CSE seems doubtable.

\begin{figure}[htbp]
\includegraphics
[width=8.6cm]{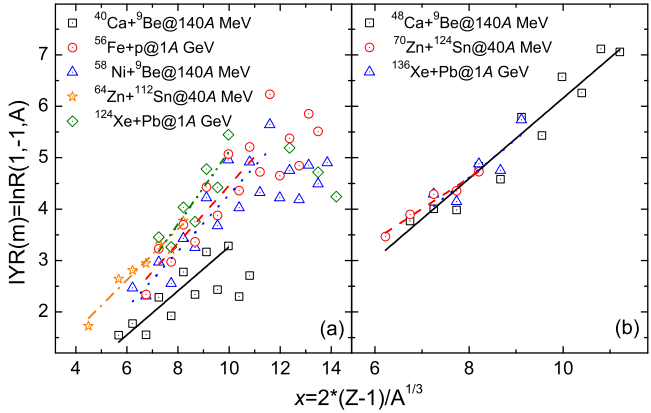}\caption{\label{lnRexp} (Color
online) The isobaric yield ratio for mirror nuclei [IYR(m)] in measured
reactions. The projectile nuclei are n/p symmetric and neutron-rich
in (a) and (b), respectively. The lines denote the
linear fitting results of IYR(m).}
\end{figure}

To clarify the possibility of the isospin dependence of IYR(m), in Fig. \ref{lnRexp}
some IYR(m) for the measured reactions are plotted. The used reactions are the 140$A$
MeV $^{40,48}$Ca + $^{9}$Be \cite{Mocko06}, 1$A$ GeV $^{124,136}$Xe + Pb \cite{Henz08},
1$A$ GeV $^{56}$Fe + $p$ \cite{CantonFe07}, 40$A$ MeV $^{64}$Zn + $^{112}$Sn and
$^{70}$Zn + $^{124}$Sn \cite{HuangRoy}. In (a) of Fig. \ref{lnRexp}, IYR(m) are for
reactions using the neutron-proton symmetric projectiles, which show similar distributions
except the $^{40}$Ca reaction. When $x>11$, these IYR(m) reach plateaus. The
IYR(m) for the $^{64}$Zn reaction is different to others maybe due to the projectile-like
fragments are removed in the measurements. The IYR(m) for the $^{124}$Xe reaction is
plotted in (a) since $^{124}$Xe is $N/Z$ symmetric comparing to $^{136}$Xe, and actually
the IYR(m) for the $^{124}$Xe reaction does show the character of IYR(m) for those of the
$N/Z$ symmetric nuclei reactions. The overlapping IYR(m) in (b) of Fig. \ref{lnRexp} are for the
reactions using the neutron-rich projectiles. IYR(m) for these reactions in Fig. \ref{lnRexp}
indicate that the finite size effect and the volume effect are negligible in the neutron-rich
nucleus induced reaction.

The statistical abrasion-ablation (SAA) model well reproduces the yields of fragments
in HIC induced by the neutron-rich nucleus \cite{SAABrohm94,SAAGaim91,Fang07JPG,MaCW09PRC,FangPRC00,WHL10,MaCW08CPB},
and it is applied to study the isospin effects in reactions induced by the neutron-rich projectile 
\cite{MaCW09PRC,FangPRC00,MaCW09CPB,MACW10JPG,MaCW09PRC,MaCW10PRC,Fang10PRC}. It can
well reproduce the yields of both the small and large mass fragments in the 140$A$ MeV
$^{40,48}$Ca + $^9$Be and $^{58,64}$Ni + $^9$Be projectile fragmentation, and the
density effect in the fragment yield is investigated \cite{MaCW09PRC,MaCW09CPB}.
Considering the dependence of the yields and the resultant IYR(m) on the isospin of the
projectile, the "finite-size" effects of the IYR(m) shown in SGC/CSE (which depends
on the volume of the reaction system) should be reexamined. In this article, using
the SAA model, the 140$A$ MeV $^{38\sim52}$Ca nuclei which have different isospins,
and $^{72}$Zn/$^{96}$Zr/$^{120}$Sn which have the same $N/Z$ ratio as $^{48}$Ca
(which are calculated in Ref. \cite{Souza12finite}) are re-calculated to demonstrate
the dependence of the IYR(m) on the isospin and mass of the projectile ($A_p$). For
simplification, only important formula of the SAA are listed here since the model is
well described in Refs. \cite{SAABrohm94,SAAGaim91,FangPRC00,MaCW09PRC}.

The SAA model is a two-stages model, in which the first stage describes the collisions
and determine the primary fragments, and the second stage models the deexcitation of the primary fragments.
In the colliding stage, the SAA model takes independent nucleon-nucleon collisions
for participants in an overlap zone of the two colliding nuclei and determines the
distributions of abraded neutrons and protons. The colliding nuclei are described
to be composed of parallel tubes orienting along the beam direction. For an infinitesimal
tube in the projectile, the transmission probabilities for neutrons (protons) at
a given impact parameter $\vec b$ are calculated by
\begin{equation}\label{trans}
t_k(\vec s-\vec b)=\mbox{exp}\{-[\rho{_n^T}(\vec s-\vec
b)\sigma_{nk}+\rho{_n^P}(\vec s-\vec b)\sigma_{pk}]\},
\end{equation}
where $\rho^T$ is the nuclear-density distribution of the target integrated along the
beam direction, the vectors $\vec s$ and $\vec b$ are defined in the plane perpendicular
to beam, and $\sigma_{k'k}$ is the free nucleon-nucleon reaction cross section. The
average absorbed mass in the limit of infinitesimal tubes at a given $\vec b$ is,
\begin{eqnarray}
<\Delta A(b)>=\int d^{2}s \rho_{n}^{T}(\vec s)[1-t_n(\vec s-\vec b)] \nonumber\\
+\int d^{2}s \rho_{p}^{P}(\vec s)[1-t_p(\vec s-\vec b)] .
\end{eqnarray}
The $\rho_n$ and $\rho_p$ distributions are assumed to be the Fermi-type.
The cross section of a specific isotope (primary fragment) can be calculated from
\begin{equation}\label{yieldisotope}
\sigma(\Delta N, \Delta Z)=\int d^2bP(\Delta N, b)P(\Delta Z,b),
\end{equation}
where $P(\Delta N, \mathit{b})$ and $P(\Delta Z, \mathit{b})$ are
the probability distributions for the abraded neutrons and protons
at a given impact parameter $\mathit{b}$, respectively.

The second stage of the reaction in the SAA is the evaporation of the
primary fragment \cite{SAAGaim91}, which is described by the conventional statistical
model under the assumption of thermal equilibrium. After the evaporation, the isotopic
yield (final fragment) comparable to the experimental result is obtained.

\begin{figure}[htbp]
\includegraphics
[width=7cm]{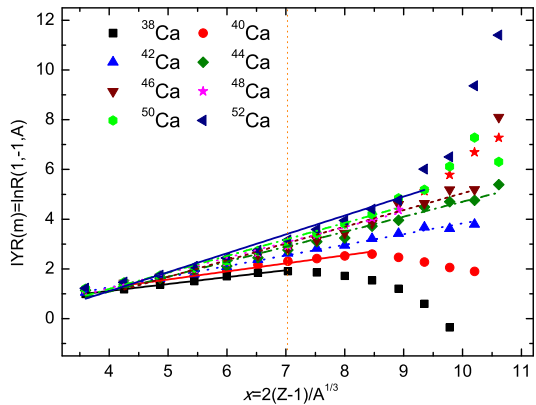}\caption{\label{CaMSAA} (Color
online) IYR(m) for the 140$A$ MeV $^{38\sim52}$Ca + $^{12}$C reactions calculated
by the SAA model. The lines denote the linear fitting results.}
\end{figure}

The IYR(m) of the calculated 140$A$ MeV $^{38\sim52}$Ca + $^{12}$C reactions are plotted
in Fig. \ref{CaMSAA}. The IYR(m) in the $^{38\sim52}$Ca reactions increases when the
projectile becomes more neutron-rich. Different to the results in SGC/CSE,
in which the linear correlation exhibits in the $x>6$ fragments, the linear correlation
exhibits in the fragments of relative small $x$. Fairly well linear correlations
between the IYR(m) and $x$ are found in the $^{42,44}$Ca reactions. For the neutron-deficient
$^{38}$Ca projectile, the IYR(m) firstly increases linearly with $x$ when $x<7$, but decreases
when $x>7$. The IYR(m) for $^{40}$Ca has a similar distribution as that of $^{38}$Ca except
the inflection $x$ is around 8.5. For the neutron-rich $^{46\sim52}$Ca projectiles, the IYR(m)
increases linearly when $x$ is not large, but it increases quickly when $x$ of fragments
is lager than 0.85$x$ of projectile. For these fragments, only few nucleons are removed
from the projectile, which are mostly produced in the peripheral collisions in the SAA
model \cite{MACW10JPG,MACW10PRC}. The yield of mirror nucleus which has small-$x$ and large mass
$x$ are affected by $\rho_n$ and $\rho_p$. The similar IYR(m) shown in the small-$x$ mirror
nucleus can be explained as the similarity of the $\rho_n$ and $\rho_p$ distributions in the core of
projectiles. The deviation of the IYR(m) shown in the large-$x$ mirror nuclei shows the difference
of $\rho_n$ and $\rho_p$ in the surface of projectile, which
can also be viewed as a skin effect of the projectile \cite{MaCW09PRC,MaCW10PRC}. The slope [only
considering the linearly increasing part of the IYR(m)] slightly increases as the neutron
numbers of the projectile increases. But for the $^{50}$Ca and $^{52}$Ca reactions, the IYR(m)
almost overlap except for some fragments with large $x$, which shows a signature of saturation
in reactions of very neutron-rich projectile.

\begin{figure}[htbp]
\includegraphics
[width=7cm]{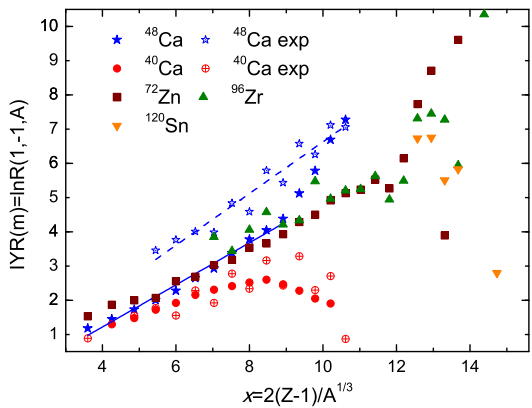}\caption{\label{CaMEXP} (Color
online) The IYR(m) in reactions calculated using SAA and measured reactions. The full
and open circles (stars) are for the SAA and experimental~\cite{Mocko06}
140$A$ MeV $^{40}$Ca($^{48}$Ca) reactions, respectively;
the triangles, hexagon, and diamonds represent the IYR(m)
for the 140$A$ MeV $^{72}$Zn/$^{96}$Zr/$^{120}$Sn + $^{12}$C reactions.
The lines are the fitting results by Eq. (\ref{lnRmirror}). }
\end{figure}

In Fig. \ref{CaMEXP}, the SAA and the experimental IYR(m) for the $^{40,48}$Ca and
the 140$A$ MeV $^{72}$Zn/$^{96}$Zr/$^{120}$Sn + $^{12}$C reactions are plotted.
Though the odd-even staggering in the experimental IYR(m) can not be reproduced, the
SAA results can fit the $^{40}$Ca experimental data rather well, including the decreasing
behavior of IYR(m) when $x>9$. For $^{48}$Ca, the calculated IYR(m) are smaller than
those of the experiments. The linear fittings of the SAA and the experimental data
read $y = (0.62\pm0.02)x-(1.24\pm0.16)$ and $y = (0.75\pm0.06 )x-(0.92\pm0.50)$,
respectively. More interesting point the figure tells us is that the large difference
among the IYR(m) in SGC/CSE in the $^{72}$Zn/$^{96}$Zr/$^{120}$Sn reactions disappears
in the SAA results. The IYR(m) for $^{72}$Zn projectile exhibits quite good linear correlation
when $x<11$, which shows very little difference from the calculated IYR(m) for $^{48}$Ca.
The mirror nuclei of small-$A$ do not survive the deexcitation process in the SAA calculation
for $^{96}$Zr and $^{120}$Sn, but the IYR(m)s for $^{72}$Zn, $^{96}$Zr, and $^{120}$Sn show
very little difference when $x<12$. Considering the saturation of the IYR(m) for the reactions
induced by neutron-rich nucleus shown in Fig. \ref{lnRexp} and \ref{CaMSAA}, the overlapping the
IYR(m) for $^{72}$Zn, $^{96}$Zr, and $^{120}$Sn reactions is reasonable.

\begin{figure}[htbp]
\includegraphics
[width=8.6cm]{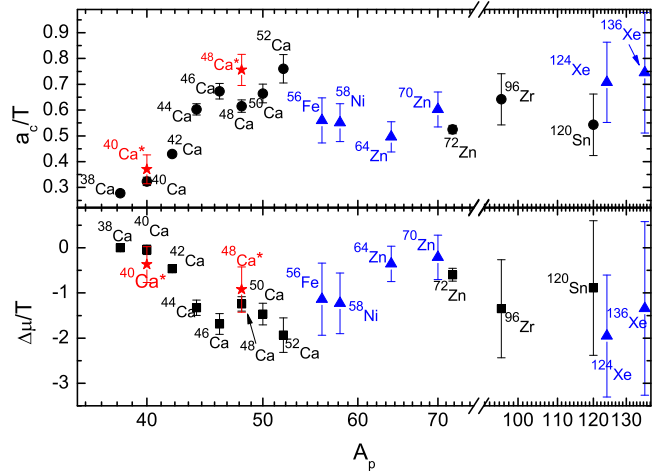}\caption{\label{acmu} (Color
online) $a_c/T$ and $\Delta\mu/T$ fitted from the IYR(m) by Eq.
(\ref{lnRmirror}). The $x$ axis represents the mass ($A_p$)
of the projectiles. The labels $^{40}$Ca*, $^{48}$Ca*, and
stars represent the values for the measured results. The
triangles represent the values for reactions shown in Fig.
\ref{lnRexp} except the Ca induced reactions.}
\end{figure}

In Fig. \ref{acmu}, $a_c/T$ and $\Delta\mu$ from the IYR(m) are plotted.
The fittings are limited to the linear part of the IYR(m). $a_c/T$ and
$\Delta\mu$ of the $^{64,70}$Zn reactions \cite{Huang10} are slightly different
to other reactions, which may due to the measurement is limited to the special
spatial angles,
and the projectile-like fragments are excluded. In the measured reactions, $a_c/T$($\Delta\mu/T$)
for the $^{56}$Fe and $^{58}$Ni reactions have very little difference, and the same
occurs in the $^{48}$Ca and $^{136}$Xe reactions. $a_c/T$ ($\Delta\mu/T$)
from the IYR(m) for the $N/Z$ symmetric projectile are smaller (larger) than those of the
neutron-rich projectile. In the SAA, the $a_c/T$ ($\Delta\mu/T$) of the IYR(m)
for the Ca-isotopes increases (decreases) when $A_p$ increases, which
shows a kind of isospin dependence. But in reactions of the large-mass projectiles,
the $a_c/T$ ($\Delta\mu/T$) obtained are very similar. $a_c/T$ and $\Delta\mu/T$
tend to be flat and saturate when the projectiles are neutron-rich (after $^{44}$Ca).
For the projectiles having the same $N/Z$, no "finite-size" effects for the larger systems
($^{72}$Zn/$^{96}$Zr/$^{120}$Sn) are shown in $a_c/T$ or $\Delta\mu/T$. It is easy
to conclude that when the projectile is not very neutron-rich, the IYR(m) depends on the
isospin of the projectile, but its size dependence cannot be excluded. When the
projectile is neutron-rich, the isospin dependence of the IYR(m)
is weakened and even disappears, and the size dependence of the IYR(m) can be excluded.

According to Eq. (\ref{yieldisotope}), the yield of a primary fragment is mainly determined
by the nucleon-nucleon cross sections, the $\rho_n$ and $\rho_p$ distributions.
Different to the strictly linear increase of the IYR(m) in the $x>6$ fragments
in SGC/CSE, in the SAA results the IYR(m) in fragments with small $x$ has very little
difference, but show sudden changes in the large-$x$ fragments. This
is also shown in the measured data in Fig. \ref{lnRexp}(a). Beside the dependence
of the projectile mass, the phenomenon can also be explained as the isospin effects,
which is similar to the isospin phenomena shown in HIC \cite{MaCW09PRC,Fang07JPG,MaCW09CPB,MACW10JPG,MaCW11CPC,MaCW12PRCT,MYGisoeff-LG-cpl}.
The similarity in the IYR(m) of the small-$x$ fragments can be explained being the similarity of the $\rho_n$
and $\rho_p$ distributions in the central collisions, while the large difference
in the IYR(m) of the large-$x$ fragments corresponds to the large difference between the $\rho_n$
and $\rho_p$ distributions in the peripheral collisions according to the skin region of the neutron-rich
projectile \cite{MaCW09PRC,MaCW10PRC}. It is concluded that both the isospin effects
and the size effects are obvious for the not very neutron-rich reactions. But for the very
neutron-rich projectile, contrary conclusions to the "finite-size" effects in SGC/CSE
are found, i.e., the dependence of the IYR(m) on the system size disappears.

Using one reaction, the extracted $(\mu_n-\mu_p)/T$ and $a_{c}/T$ by the IYR method could
depend on the $N/Z$ ratio of projectile. In Ref.~\cite{Huang10}, the $a_{c}/T$ is scaled
to the reaction system parameter Z/A($\equiv (Z_p+Z_t)/(A_p+A_t)$). If using Huang's methods
\cite{Huang10} to extract the $a_{sym}/T$ of the fragments, an increase of 0.15 of
$a_{c}/T$ may introduce a difference of 0.6$\sim$1.5 when $x$ increases from 4 to 10,
i.e., for the small-$x$ fragment, the difference is rather small. Based on these results,
it is suggested that the isobaric yield ratio only has little "finite-size" effects, especially in
reactions of very neutron-rich projectile.

To summarize, the "finite-size" effects in the IYR(m) in the SGC/CSE models are reexamined using
a modified SAA model by considering the influence of density in the yield of the fragment in HICs.
The SAA and the experimental results reveal that, when the projectile is not so neutron-rich,
the IYR(m) depends on the isospin of projectile nucleus, but the size dependence of
the IYR(m) can not be excluded. When the projectile is very neutron-rich, the IYR(m)
dependence of the isospin of projectile is weakened and disappears, and the IYR(m)
does not depend on the mass (or volume) of the projectile. The "finite-size" effects shown in
SGC/CSE, which depend on the mass or the volume of the projectile, disappear in the
reactions induced by the neutron-rich projectile both in the SAA and experimental
results.

\textit{This work is supported by the National Natural Science
Foundation of China under contract Nos. 10905017 and 10979074,
the Knowledge Innovation Project of the Chinese Academy of Sciences under contract
No. KJCX2-EW-N01, Program for Science\&Technology Innovation Talents in Universities
of Henan Province (HASTIT), and the Young Teacher Project in Henan Normal University, China.  We thank Dr. Mei-Rong Huang and Professor Roy Wada at the Institute of Modern Physics,
Chinese Academy of Science of China for providing us the experimental results
of the $^{64,72}$Zn + $^{124}$Sn reactions.}

\end{document}